\documentclass[10pt,conference]{IEEEtran}
\IEEEoverridecommandlockouts\IEEEpubid{\makebox[\columnwidth]{ 978-1-6654-5975- 4/22~\copyright~2022 IEEE \hfill} \hspace{\columnsep}\makebox[\columnwidth]{ }}
\usepackage{amsmath,amssymb}
\usepackage{graphicx,graphics,color,psfrag}
\usepackage{cite,balance}
\usepackage{algorithm}
\usepackage{accents}
\usepackage{bm}
\usepackage{url}
\usepackage{algorithmic}
\usepackage[english]{babel}
\usepackage{multirow}
\usepackage{enumerate}
\usepackage{cases}
\usepackage{stfloats}
\usepackage{dsfont}
\usepackage{color,soul}
\usepackage{amsfonts}
\usepackage{tcolorbox}
\usepackage{amsmath}
\usepackage{float}

\usepackage{cite,graphicx,amsmath,amssymb}
\usepackage{fancyhdr}
\usepackage{hhline}
\usepackage{graphicx,graphics}
\usepackage{array,color}
\usepackage{amsmath}
\usepackage{stfloats}
\usepackage[flushleft]{threeparttable}

\newtheorem{proposition}{Proposition}

\newtheorem{lemma}{Lemma}

\ifCLASSINFOpdf

\else
 
\fi

\begin{document}

\title{Intelligent Reflecting Surface Enabled Sensing: Cram\'er-Rao Lower Bound Optimization}

\author{
\IEEEauthorblockN{Xianxin~Song\IEEEauthorrefmark{1},  Jie~Xu\IEEEauthorrefmark{1}, Fan~Liu\IEEEauthorrefmark{2}, Tony~Xiao~Han\IEEEauthorrefmark{3}, and  Yonina C.~Eldar\IEEEauthorrefmark{4}
\thanks{The work was supported in part by the National Key R\&D Program of China with grant No.2018YFB1800800, the Basic Research Project No. HZQB-KCZYZ-2021067 of Hetao Shenzhen-HK S\&T Cooperation Zone, the National Natural Science Foundation of China under grants No. U2001208, 62101234 and U20B2039, the Science and Technology Program of Guangdong Province under grant No. 2021A0505030002, the Shenzhen Fundamental Research Program under grant No. 20210318123512002, the Guangdong Provincial Key Laboratory of Future Networks of Intelligence under grant No. 2022B1212010001, the European Research Council (ERC) through the European Union’s Horizon 2020 research and innovation programme under grant No. 101000967. J. Xu is the corresponding author. }%
}

\IEEEauthorblockA{\IEEEauthorrefmark{1}School of Science and Engineering, Future Network of Intelligence Institute, and Guangdong Provincial Key \\ Laboratory of Future Networks of Intelligence, The Chinese University of Hong Kong (Shenzhen),  China}
\IEEEauthorblockA{\IEEEauthorrefmark{2}Department of Electrical and Electronic Engineering, Southern University of Science and Technology, China}
\IEEEauthorblockA{\IEEEauthorrefmark{3}Wireless Technology Lab, 2012 Laboratories, Huawei, China}
\IEEEauthorblockA{\IEEEauthorrefmark{4}Faculty of Mathematics and Computer Science, Weizmann Institute of Science, Israel}
Email: xianxinsong@link.cuhk.edu.cn,  xujie@cuhk.edu.cn, liuf6@sustech.edu.cn,
\\  tony.hanxiao@huawei.com, yonina.eldar@weizmann.ac.il% <-this % stops a space
}

\maketitle

\begin{abstract}
This paper investigates intelligent reflecting surface (IRS) enabled non-line-of-sight (NLoS) wireless sensing, in which an IRS is deployed to assist an access point (AP) to sense a target in its NLoS region. It is assumed that the AP is equipped with multiple antennas and the IRS is equipped with a uniform linear array. The AP aims to estimate the target's direction-of-arrival (DoA) with respect to the IRS, based on the echo signals from the AP-IRS-target-IRS-AP link. Under this setup, we jointly design the transmit beamforming at the AP and the reflective beamforming at the IRS to minimize the Cram\'er-Rao lower bound (CRLB) on estimation error. Towards this end, we first obtain the CRLB expression for estimating the DoA in closed form. Next, we optimize the joint beamforming design to minimize the CRLB, via alternating optimization, semi-definite relaxation, and successive convex approximation. Numerical results show that the proposed design based on CRLB minimization achieves improved sensing performance in terms of mean squared error, as compared to the traditional schemes with signal-to-noise ratio maximization and separate beamforming.
\end{abstract}

%\begin{IEEEkeywords}
%Intelligent reflecting surface, non-line-of-sight wireless sensing, Cram\'er-Rao lower bound, joint transmit and reflective beamforming.
%\end{IEEEkeywords}

\IEEEpeerreviewmaketitle
%\addtolength{\topmargin}{-0.04in}
\section{Introduction}
Integrating wireless sensing into future beyond-fifth generation (B5G) and six-generation (6G) wireless networks as a new functionality has attracted growing research interest to support future industrial Internet of things (IoT) applications with environment awareness (see, e.g., \cite{liu2020joint,liu2021integrated} and the references therein). Conventionally, wireless sensing relies on line-of-sight (LoS) links between access points (APs) and sensing targets, such that the sensing information is extracted based on the target echo signals. However, in practical scenarios,  sensing targets are likely to be located at the non-LoS (NLoS) region of APs, where conventional LoS sensing is not applicable in general. Therefore, how to realize NLoS sensing in such scenarios is a challenging task. 

Motivated by its success in wireless communications \cite{8811733,9122596}, intelligent reflecting surface (IRS) or reconfigurable intelligent surface (RIS) have become a viable solution to realize NLoS wireless sensing (see, e.g., \cite{aubry2021reconfigurable,shao2022target,9361184,Stefano,song2021joint,9364358}). By properly deploying IRSs around the AP to reconfigure the radio propagation environment, virtual LoS links can be established between the AP and the targets in its NLoS region, such that the AP can perform NLoS target sensing based on the echo signals from  AP-IRS-target-IRS-AP links. To combat the severe signal propagation loss over such triple-reflected links, IRS adaptively control the phases at reflecting elements, such that the reflected signals are beamed towards desired directions to enhance sensing performance.

There have been several prior works investigating IRS-enabled wireless sensing \cite{aubry2021reconfigurable,shao2022target,9361184,Stefano} and IRS-enabled integrated sensing and communications (ISAC) \cite{song2021joint,9364358}, respectively. The work \cite{aubry2021reconfigurable} presented the NLoS radar equation based on the AP-IRS-target-IRS-AP link and evaluated the resultant sensing performance in terms of signal-to-noise ratio (SNR) and signal-to-clutter ratio (SCR). In \cite{shao2022target}, the authors considered an IRS-enabled bi-static target estimation, where dedicated sensors were installed at the IRS for estimating the direction of its nearby target through the IRS controller-IRS-target-sensors and IRS controller-target-sensors links. 
%Under this setup, the authors optimized the IRS's reflective beamforming to maximize the average received signal power at the sensors and applied the multiple signal classification (MUSIC) algorithm for estimating targets.
In \cite{9361184} and \cite{Stefano}, the authors considered IRS-enabled target detection, in which the IRS's passive beamforming was optimized to maximize the target detection probability subject to a fixed false alarm probability. In \cite{song2021joint}, the authors considered an IRS-enabled ISAC system with one base station (BS), one communication user (CU), and multiple targets, in which the IRS’s minimum beampattern gain towards the desired sensing angles was maximized by jointly optimizing the transmit and reflective beamforming, subject to the minimum SNR requirement at the CU.
In \cite{9364358}, the authors considered an IRS-enabled ISAC system with one BS, one CU, and one target, in which the SNR of radar was maximized by joint beamforming design while ensuring the SNR at the CU.

In prior works on IRS-enabled sensing and IRS-enabled ISAC, the sensing SNR (or beampattern gain) and the target detection probability have been widely adopted as the sensing performance measure. Cram\'er-Rao lower bound (CRLB) is another important sensing performance measure, especially for target estimation, which provides a lower bound on the variance of unbiased parameter estimators. In prior studies on wireless sensing\cite{kay1993fundamentals,bekkerman2006target} and ISAC\cite{9652071} without IRS, CRLB has been widely adopted as the design objective for sensing performance optimization. Nevertheless, to our best knowledge, how to analyze the CRLB performance for NLoS target estimation through the AP-IRS-target-IRS-AP link and optimize such performance by joint transmit and reflective beamforming design has not been explored. 

This paper considers an IRS-enabled NLoS wireless sensing system, in which the AP aims to estimate the target's direction-of-arrival (DoA) with respect to (w.r.t.) the IRS based on the echo signals from the AP-IRS-target-IRS-AP link. 
%In particular, we focus on the narrowband transmission for IRS-enabled sensing, in which a general multi-path channel model (including both LoS and NLoS paths) is considered for the AP-IRS link, and an LoS channel model is considered for the IRS-target link to facilitate DoA estimation. 
%It is  assumed that the AP perfectly knows the channel state information (CSI) of the AP-IRS link, and needs to estimate the target's DoA.
We aim to minimize the CRLB for DoA estimation, by jointly optimizing the transmit beamforming at the AP and the reflective beamforming at the IRS. Towards this end, we first obtain the closed-form CRLB expression for estimating the DoA. Based on the obtained CRLB, it is shown that the target's DoA is only estimable when the rank of the AP-IRS channel matrix is larger than one or equivalently there are more than one signal paths in that channel. Next, we minimize the obtained CRLB by joint beamforming design, subject to a maximum power constraint at the AP. 

We present an efficient algorithm for CRLB minimization via alternating optimization, semi-definite relaxation (SDR), and successive convex approximation (SCA). Then, we present the maximum likelihood estimation (MLE) to estimate the target's DoA, for which the achieved estimation mean squared error (MSE) is shown to coverge towards the CRLB when the SNR is sufficiently high. Finally, numerical results show that the proposed joint beamforming design based on CRLB minimization achieves improved sensing performance in terms of estimation MSE, as compared to the traditional schemes with SNR maximization and separate beamforming designs.

\textit{Notations:} 
Boldface letters refer to vectors (lower case) or matrices (upper case). For a square matrix $\mathbf S$, $\mathbf S^{-1}$ denotes its inverse, and $\mathbf S \succeq \mathbf{0}$ means that $\mathbf S$ is positive semi-definite. For an arbitrary-size matrix $\mathbf M$, $\mathrm {rank}(\mathbf M)$, $\mathbf M^*$, $\mathbf M^{\mathrm {T}}$, and $\mathbf M^{H}$ are its rank, conjugate, transpose, and conjugate transpose, respectively. The matrix $\mathbf I_m$ is an identity matrix of dimension $m$. We use $\mathcal{C N}(\mathbf{x}, \mathbf{\Sigma})$ to denote the distribution of a circularly symmetric complex Gaussian (CSCG) random vector with mean vector $\mathbf x$ and covariance matrix $\mathbf \Sigma$, and $\sim$ to denote “distributed as”. The spaces of $x \times y$ complex and real matrices are denoted by $\mathbb{C}^{x \times y}$ and  $\mathbb{R}^{x \times y}$, respectively. The real and imaginary parts of a complex number are denoted by $\mathrm{Re}\{\cdot\}$ and $\mathrm{Im}\{\cdot\}$, respectively. The imaginary unit of a complex number is denoted by $\jmath=\sqrt{-1}$.
The symbol $\|\cdot\|$ stands for the Euclidean norm, $|\cdot|$ for the magnitude of a complex number, $\mathrm {diag}(a_1,\cdots,a_N)$ for a diagonal matrix with diagonal elements $a_1,\cdots,a_N$, 
%$\otimes$ for the Kronecker product,  
$\mathrm{vec}(\cdot)$ for the vectorization operator, and $\mathrm {arg}(\mathbf x)$ for a vector with each element being the phase of the corresponding element in $\mathbf x$. 
\begin{figure}[t]
\centering
\includegraphics[scale = 0.72]{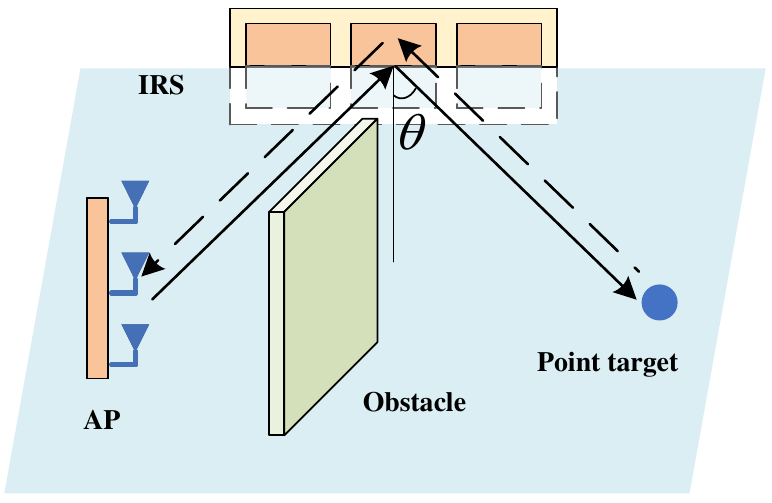}
\caption{System model of IRS-enabled sensing.}
\label{system_model}
\end{figure}

%\addtolength{\topmargin}{-0.09in}
\section{System Model}

We consider an IRS-enabled NLoS wireless sensing system as shown in Fig.~\ref{system_model}, which consists of one AP with $M>1$ antennas, one uniform linear array (ULA) IRS with $N>1$ reflecting elements, and one target at the NLoS region of the AP. The IRS is deployed to create a virtual LoS link to facilitate target sensing. In this case, the AP transmits a sensing signal and then estimates the target's DoA w.r.t. the IRS based on the echo signals from the AP-IRS-target-IRS-AP link. Target estimation is implemented at the AP, as the IRS is assumed to be a passive device without the capability of signal processing.

First, we consider transmit and reflective beamforming at the AP and the IRS, respectively. Let $\mathbf x(t)\in \mathbb{C}^{M \times 1}$ denote the transmitted signal by the AP at time slot $t$ and $T$ the radar dwell time. The sample covariance matrix of the transmitted signal is 
$\mathbf R_x=\frac{1}{T}\sum_{t=1}^T \mathbf x(t)\mathbf x(t)^{H}$.
%, which corresponds to the transmit beamforming vectors to be optimized.
Here, $\mathrm {rank}(\mathbf R_x)$ corresponds to the number of sensing beams sent by the AP, each of which can be obtained via the eigenvalue decomposition (EVD) of $\mathbf R_x$\cite{hua}.
%Notice that in order to provide more degrees of freedom for sensing, the AP transmits multiple sensing beams, i.e., $0 \le \mathrm {rank}(\mathbf R_x) \le M$ \cite{hua}.
%Suppose that $\mathrm{rank}(\mathbf R_x) = k$ and the eigenvalue decomposition (EVD) of $\mathbf R_x$ is given by $\mathbf R_x=\mathbf W \mathbf \Lambda \mathbf W^{H}$, where $\mathbf \Lambda = \mathrm{diag}(\lambda_1, \cdots, \lambda_M)$ and $\mathbf W = [\mathbf w_1,\cdots ,\mathbf w_M] $ with $\lambda_1 \ge \cdots \ge \lambda_k > \lambda_{k+1}= \cdots =\lambda_M= 0$, and $\mathbf W  \mathbf W^{H} = \mathbf W^{H} \mathbf W= \mathbf{I}_M$. This means that there are a number of $k$ sensing beams transmitted by the AP, each of which is denoted by $\sqrt{\lambda_i}\mathbf w_i, i=1, \cdots, k$ (see, e.g., \cite{hua}). 
We assume that the IRS can only adjust the phase shifts of its reflecting elements \cite{8811733}. Let $\mathbf v = [e^{\jmath \phi_1},\cdots,e^{\jmath \phi_{N}}]^{T}$ denote the reflective beamforming vector at the IRS, with $\phi_n \in (0, 2\pi]$ being the phase shift of element $n \in \{1,\cdots,N\}$, which can be optimized to enhance the sensing performance.

Next, we introduce the channel models. We focus on narrowband transmission. We consider a general multi-path model for the AP-IRS link. Accordingly, let $\mathbf{G} \in \mathbb{C}^{N \times M}$ denote the associated channel matrix of the AP-IRS link, where $\mathrm{rank}(\mathbf{G}) \ge 1$ holds in general. We consider an LoS model for the IRS-target link to facilitate DoA estimation. Let $\theta$ denote the target's DoA w.r.t. the IRS. Accordingly, the steering vector at the IRS with angle $\theta$ is 
\begin{equation}\label{eq:steering_vector}
\mathbf a(\theta) = [1,e^{\jmath 2\pi\frac{d_\text{IRS}\sin \theta}{\lambda_\text{IRS}}},\cdots,e^{\jmath 2\pi\frac{ (N-1)d_\text{IRS}\sin \theta}{\lambda_\text{IRS}}}]^{T},
\end{equation}
where $d_\text{IRS}$ denotes the spacing between consecutive reflecting elements at the IRS and $\lambda_\text{IRS}$ denotes the carrier wavelength. The target response matrix w.r.t. the IRS (or equivalently the cascaded IRS-target-IRS channel) is 
$\mathbf H = \alpha\mathbf a(\theta)\mathbf a^{T} (\theta)$,
where $\alpha \in \mathbb{C}$ denotes the complex-valued channel coefficient dependent on the target's radar cross section (RCS) and the round-trip path loss of the IRS-target-IRS link. 
%Note that the target's DoA $\theta$ and the channel coefficient $\alpha$ are both unknown parameters at the AP. 

Based on the transmitted signal $\mathbf x(t)$ and channel models, the signal impinged at the IRS is $\mathbf G \mathbf x(t)$. After the reflective beamforming at the IRS and target reflection, the echo signal impinged at the IRS becomes $\mathbf H \mathbf \Phi \mathbf G \mathbf x(t)$, where $\mathbf{\Phi} = \mathrm {diag}(\mathbf v)$ denotes the reflection matrix of the IRS. By further reflective beamforming at IRS and through the IRS-AP channel, the received echo signal at the AP through the AP-IRS-target-IRS-AP link at time $t\in\{1,\cdots, T\}$ is
\begin{equation}\label{eq:echo_signal}
\begin{split}
\mathbf y(t)=&\mathbf{G}^{T}\mathbf{\Phi}^{T}\mathbf H \mathbf{\Phi}{\mathbf{G}}\mathbf x(t)+ \mathbf n(t)\\
=&\alpha\mathbf{G}^{T}\mathbf{\Phi}^{T}\mathbf a(\theta)\mathbf a^{T} (\theta)\mathbf{\Phi}{\mathbf{G}}\mathbf x(t)+ \mathbf n(t),  
\end{split}
\end{equation}
where $\mathbf n(t) \sim \mathcal{C N}(\mathbf{0}, \sigma_\text{R}^2\mathbf I_M)$ denotes the additive white Gaussian noise (AWGN) at the AP receiver.  

The AP aims to estimate the target's DoA $\theta$, based on the received echo signal in \eqref{eq:echo_signal}. It is assumed that the AP perfectly knows the channel state information (CSI) $\mathbf G$ of the AP-IRS link via proper channel estimation algorithms (see, e.g., \cite{9722893}). 
%This assumption is practically valid, which is due to the fact that the AP and the IRS are deployed at fixed locations and thus their channels are slowly varying in practice. 
%It is also assumed that the AP perfectly knows its sample covariance matrix $\mathbf R_x$ and the IRS's reflective beamforming vector $\mathbf v$.
%It is also assumed that the AP perfectly knows its transmitted signal $\mathbf x(t)$ (and the associated sample covariance matrix $\mathbf R_x$) and the IRS's reflective beamforming vector $\mathbf v$, which can be optimized to enhance the sensing performance. 

%\addtolength{\topmargin}{-0.16in}

\section{Estimation CRLB Derivation}
This section analyzes the estimation performance in terms of CRLB. Let $\boldsymbol \xi=[\theta,  \tilde{\boldsymbol\alpha}^{T}]^{T} \in \mathbb{R}^{3\times 1}$ denote the vector of unknown parameters to be estimated, including the target's DoA $\theta$ and the complex-valued channel coefficient $\alpha$, where $\tilde{\boldsymbol\alpha}=[\mathrm{Re}\{\alpha\},\mathrm{Im}\{\alpha\}]^{T}$.  We are particularly interested in characterizing the CRLB for DoA estimation. This is due to the fact that it is difficult to extract the target information from the channel coefficient $\alpha$, as it depends on both the target's RCS and the round-trip path loss that are usually unknown.

\setcounter{equation}{10} 
\newcounter{TempEqCnt}
\begin{figure*}[ht]
\begin{equation}\label{eq:CRB_1}
\mathrm{CRLB}(\theta)=\frac{\sigma_\text{R}^2 \lambda_\text{IRS}^2}{ 8T|\alpha|^2\pi^2d_\text{IRS}^2\cos^2(\theta)\left(\mathbf v^{H} \mathbf R_2 \mathbf v \left(\mathbf v^{H} \mathbf D \mathbf R_1 \mathbf D\mathbf v-\frac{|\mathbf v^{H} \mathbf D \mathbf R_1 \mathbf v|^2}{\mathbf v^{H} \mathbf R_1 \mathbf v}\right) + \mathbf v^{H} \mathbf R_1 \mathbf v \left(\mathbf v^{H} \mathbf D \mathbf R_2 \mathbf D \mathbf v-\frac{|\mathbf v^{H} \mathbf D \mathbf R_2 \mathbf v|^2}{\mathbf v^{H} \mathbf R_2 \mathbf v}\right)\right)}
\end{equation}
\hrulefill
\end{figure*}

First, we obtain the Fisher information matrix (FIM) for estimating $\boldsymbol \xi$ to facilitate the derivation of the CRLB for DoA estimation. Towards this end, we stack the transmitted signals, the received signals, and the noise over the radar dwell time as $\mathbf X =[\mathbf x(1),\cdots,\mathbf x(T)]$, $\mathbf Y =[\mathbf y(1),\cdots,\mathbf y(T)]$, and $\mathbf N =[\mathbf n(1),\cdots,\mathbf n(T)]$, respectively. Accordingly, we have 
\setcounter{equation}{2} 
\begin{equation}\label{eq: Y}
\mathbf Y = \alpha\mathbf B(\theta)\mathbf X+\mathbf N,
\end{equation}
where $\mathbf B(\theta)= \mathbf b(\theta)\mathbf b(\theta)^{T}$ with $\mathbf b(\theta)= \mathbf{G}^{T}\mathbf{\Phi}^{T}\mathbf a(\theta)$. For notational convenience, in the sequel we drop $\theta$ in $\mathbf a(\theta)$, $\mathbf b(\theta)$, and $\mathbf B(\theta)$, and accordingly denote them as $\mathbf a$, $\mathbf b$, and $\mathbf B$, respectively. By vectorizing \eqref{eq: Y}, we have
\begin{equation}\label{eq:vec_data}
\tilde{\mathbf y}=\mathrm{vec}(\mathbf Y)=\tilde{\mathbf u} + \tilde{\mathbf n},
\end{equation}
where $\tilde{\mathbf u} = \alpha \mathrm{vec}(\mathbf B \mathbf X)$ and $\tilde{\mathbf n}=\mathrm{vec}(\mathbf N) \sim \mathcal{C N}(\mathbf{0}, \mathbf{R}_n)$ with $\mathbf R_n=\sigma_\text{R}^2\mathbf I_{MT}$. 
Let $\mathbf F  \in \mathbb{R}^{3 \times 3}$ denote the FIM matrix for estimating $\boldsymbol\xi$ based on \eqref{eq:vec_data}. Each element of $\mathbf F$ is given by \cite{kay1993fundamentals}
\begin{equation}\label{eq:FIM}
\mathbf F_{i,k}=\frac{2}{\sigma^2}\mathrm{Re}\left\{\frac{\partial \tilde{\mathbf u}^{H}}{\partial \boldsymbol\xi_i}\frac{\partial \tilde{\mathbf u}}{\partial \boldsymbol\xi_k}\right\}, i,k=1,2,3.
\end{equation}
Based on \eqref{eq:FIM}, the FIM matrix $\mathbf F$ can be partitioned as 
\begin{equation}\label{eq:FIM_partitioned}
\mathbf F=
\begin{bmatrix}
\mathbf{F}_{\theta \theta} & \mathbf{F}_{\theta \tilde{\boldsymbol\alpha}}\\
\mathbf{F}^{T}_{\theta \tilde{\boldsymbol\alpha}} & \mathbf{F}_{\tilde{\boldsymbol\alpha} \tilde{\boldsymbol\alpha}}
\end{bmatrix},
\end{equation}
where
$\mathbf F_{\theta \theta}=\frac{2T|\alpha|^2}{\sigma_\text{R}^2}\text{tr}(\dot {\mathbf B}  \mathbf R_x \dot {\mathbf B}^{H})$,
$\mathbf F_{\theta \tilde{\boldsymbol\alpha}}=\frac{2T}{\sigma_\text{R}^2}\mathrm{Re}\{\alpha^*\text{tr}( {\mathbf B}\mathbf R_x \dot {\mathbf B}^{H} )[1,\jmath]\}$, and
$\mathbf F_{\tilde{\boldsymbol\alpha} \tilde{\boldsymbol\alpha}}=\frac{2T}{\sigma_\text{R}^2}\text{tr}( {\mathbf B} \mathbf R_x  {\mathbf B^{H}} )\mathbf I_2$,
with $\dot{\mathbf B} =\frac{\partial \mathbf B }{\partial \theta}$ denoting the partial derivative of $\mathbf B$ w.r.t. $\theta$.  The derivation of the FIM follows the standard procedure in \cite{bekkerman2006target}; see details in the extended version of this paper \cite[Appendix A]{xianxin}. 

Next, we derive the CRLB for estimating the DoA, which corresponds to the first diagonal element of $\mathbf F^{-1}$, i.e., 
\begin{equation}\label{eq:FIM_the}
\mathrm{CRLB}(\theta) =[\mathbf F^{-1}]_{1,1} =[\mathbf F_{\theta \theta}-\mathbf F_{\theta \tilde{\boldsymbol\alpha}}\mathbf F_{\tilde{\boldsymbol\alpha} \tilde{\boldsymbol\alpha}}\mathbf F_{\theta \tilde{\boldsymbol\alpha}}^{T}]^{-1}.
\end{equation}
Based on  \eqref{eq:FIM_the} and \eqref{eq:FIM_partitioned}, we have the following lemma.
\begin{lemma} \label{lemma1}
The CRLB for estimating the DoA $\theta$ is given by 
\begin{equation}\label{eq:CRB}
\mathrm{CRLB}(\theta)=\frac{\sigma_\text{R}^2}{2T|\alpha|^2\left(\mathrm{tr}(\dot {\mathbf B} \mathbf R_x \dot {\mathbf B}^{H})-\frac{|\mathrm{tr}(\mathbf B \mathbf R_x  \dot {\mathbf B}^{H})|^2}{\mathrm{tr}(\mathbf B \mathbf R_x \mathbf B^{H})}\right)}.
\end{equation}
\end{lemma} 

To gain more insight and facilitate the reflective beamforming design, we re-express $\mathrm{CRLB}(\theta)$ in \eqref{eq:CRB} w.r.t. the reflective beamforming vector $\mathbf v$. Towards this end, we introduce $\mathbf A=\mathrm{diag}(\mathbf a)$, and accordingly have $\mathbf b=\mathbf{G}^{T}\mathbf{\Phi}^{T}\mathbf a = \mathbf{G}^{T}\mathbf A\mathbf{v}$. Then, let $\dot{\mathbf b}$ denote the partial derivative of $\mathbf b$ w.r.t. $\theta$, where $\dot{\mathbf b}=\jmath 2\pi \frac{d_\text{IRS}}{\lambda_\text{IRS}} \cos\theta\mathbf{G}^{T}\mathbf{\Phi}^{T}\mathbf D\mathbf a= \jmath 2\pi \frac{d_\text{IRS}}{\lambda_\text{IRS}} \cos\theta\mathbf{G}^{T}\mathbf A\mathbf D\mathbf{v}$ with $\mathbf D = \mathrm{diag}(0,\cdots,N-1)$. As a result, 
\begin{equation}\label{eq:B}
\mathbf B  
= \mathbf b \mathbf b ^{T}
= \mathbf{G}^{T}\mathbf A\mathbf{v} \mathbf v^{T}\mathbf A^{T}\mathbf G,
\end{equation}
\begin{equation}\label{eq:B_dot}
\begin{split}
\dot {\mathbf B}  
=& \dot {\mathbf b} \mathbf b^{T} +\mathbf b \dot {\mathbf b}^{T}\\
=&\jmath 2\pi \frac{d_\text{IRS}}{\lambda_\text{IRS}} \cos\theta \mathbf{G}^{T}\mathbf A(\mathbf D\mathbf{v} \mathbf v^{T}+ \mathbf{v} \mathbf v^{T}\mathbf D^{T})\mathbf A^{T}\mathbf G.
\end{split}
\end{equation}
Substituting \eqref{eq:B} and \eqref{eq:B_dot} into \eqref{eq:CRB}, we re-express $\mathrm{CRLB}(\theta)$  w.r.t. $\mathbf v$ in \eqref{eq:CRB_1} at the top of this page, where $\mathbf R_1=\mathbf A^{H}\mathbf{G}^*\mathbf{G}^{T}\mathbf A$ and $\mathbf R_2=\mathbf A^{H}  \mathbf{G}^* \mathbf R^*_x \mathbf{G}^{T}\mathbf A$.

We then have the following proposition.

\begin{proposition} \label{proposition1}
If $\mathrm{rank}(\mathbf G)= 1$ (or equivalently there is only one single path between the AP and the IRS), then the FIM $\tilde{\mathbf F}$ in \eqref{eq:FIM} for estimating $\bm \xi$ is a singular matrix, and $\mathrm{CRLB}(\theta) = \infty$. Otherwise, the FIM $\tilde{\mathbf F}$ is invertible, and $\mathrm{CRLB}(\theta)$ is bounded.
\end{proposition} 
\begin{IEEEproof}
Refer to the extended version of this paper \cite[Appendix B]{xianxin}.
\end{IEEEproof}

Proposition \ref{proposition1} shows that the target's DoA $\theta$ is estimable only when $\mathrm{rank}(\mathbf G)>1$ (i.e., the number of signal paths between the AP and the IRS is larger than one). The reasons are intuitively explained as follows.  When $\mathrm{rank}(\mathbf G)= 1$, the truncated singular value decomposition (SVD) of $\mathbf G$ is expressed as $\mathbf G= \sigma_1 \hat{\mathbf u}_1 \hat{\mathbf v}_1^{T}$, where $\hat{\mathbf u}_1$ and $\hat{\mathbf v}_1$ are the left and right dominant singular vectors, and $\sigma_1$ denotes the dominant singular value. The received echo signal in \eqref{eq:echo_signal} at the AP is 
\setcounter{equation}{11} 
\begin{equation}\label{eq:echo_signal_G}
\mathbf y(t)= \alpha\sigma_1^2 \hat{\mathbf v}_1 \underbrace{\hat{\mathbf u}_1^{T}\mathbf {\mathbf{\Phi}}^{T}\mathbf a\mathbf a^{T} \mathbf{\Phi}\hat{\mathbf u}_1}_{\beta(\theta)} \underbrace{\hat{\mathbf v}_1^{T}\mathbf x(t)}_{\tilde x(t)}+ \mathbf n(t),
\end{equation}
where $\beta(\theta) =  |\hat{\mathbf u}_1^{T}\mathbf {\mathbf{\Phi}}^{T}\mathbf a|^2$ is the only scalar term related to $\theta$. Notice that in \eqref{eq:echo_signal_G}, the complex numbers $\alpha$ and $\beta(\theta)$ are coupled, and from $\mathbf y(t)$ we can only recover one observation on $\alpha \beta(\theta)$, which is not sufficient to extract $\beta(\theta)$. \footnote{In the case with $\mathrm{rank}(\mathbf G)=1$, we can use  detection methods based on, e.g., generalized likelihood ratio test \cite{Stefano} and beam scanning \cite{aubry2021reconfigurable} to detect the existence of a target based on $\alpha \beta (\theta)$. However, these methods cannot extract the target angle $\theta$, as shown in Proposition \ref{proposition1}.}

\section{Joint beamforming design for CRLB Minimization}
In this section, we propose to jointly optimize the transmit beamforming at the AP and the reflective beamforming at the IRS to minimize the CRLB for estimating the DoA in \eqref{eq:CRB} or \eqref{eq:CRB_1}, subject to the maximum transmit power constraint at the AP. In order to implement the joint beamforming design, we assume that the AP roughly knows the information of $\theta$, which is practically valid for a target tracking scenario. The channel coefficient $\alpha$ is independent of the joint beamforming design. CRLB minimization is then formulated as
\begin{subequations}
\begin{align} \notag 
  \text{(P1)}:  \ \  \min_{\mathbf R_x, \mathbf{v}} \  \  &\mathrm{CRLB}(\theta) \\ \label{eq:power}
  \text{s.t.}  \quad   &\mathrm{tr}(\mathbf R_x) \leq P_0\\\label{eq:semi}
   &\mathbf R_x \succeq \mathbf{0}\\ \label{eq:phase_1}
  & |\mathbf v_n|=1, \forall n\in \{1,\cdots,N\},
\end{align}
\end{subequations}
where $P_0$ is the maximum transmit power budget at the AP. Problem (P1) is non-convex due to the non-concavity of the objective function and the unit-modulus constraints in \eqref{eq:phase_1}. 

To approximate the non-convex problem (P1), we propose an efficient algorithm based on alternating optimization, in which the transmit beamformers $\mathbf R_x$ and the reflective beamformer $\mathbf v$ are optimized in an alternating manner. 

\subsection{Transmit Beamforming Optimization}
First, we optimize the transmit beamformers $\mathbf R_x$ in problem (P1) under any given reflective beamformer $\mathbf v$ (or $\mathbf \Phi$), in which the CRLB formula in \eqref{eq:CRB} is used. In this case, 
%minimizing $\mathrm{CRLB}(\theta)$ is equivalent to maximizing $\mathrm{tr}(\dot {\mathbf B} \mathbf R_x \dot {\mathbf B}^{H})-\frac{|\mathrm{tr}(\mathbf B\mathbf R_x  \dot {\mathbf B}^{H})|^2}{\mathrm{tr}(\mathbf B \mathbf R_x \mathbf B^{H})}$. As a result, 
the transmit beamforming optimization problem is formulated as
\begin{subequations}
\begin{align} \notag
  \text{(P2)}: \ \ \max_{\mathbf R_x} & \  \  \mathrm{tr}(\dot {\mathbf B} \mathbf R_x \dot {\mathbf B}^{H})-\frac{|\mathrm{tr}(\mathbf B\mathbf R_x  \dot {\mathbf B}^{H})|^2}{\mathrm{tr}(\mathbf B \mathbf R_x \mathbf B^{H})} \\\notag
  \text{s.t.} &  \quad \eqref{eq:power}~\text{and}  ~\eqref{eq:semi}.
\end{align}             
\end{subequations}
By introducing an auxiliary variable $t$, problem (P2) is equivalently re-expressed as 
\setcounter{equation}{13} 
\begin{subequations}
\begin{align} \notag
  \text{(P2.1)}: \ \ \max_{\mathbf R_x,t}&  \  \  t \\  \label{eq:Schur’s complement}
  \text{s.t.}&   \ \ \mathrm{tr}(\dot {\mathbf B} \mathbf R_x \dot {\mathbf B}^{H})-\frac{|\mathrm{tr}(\mathbf B\mathbf R_x  \dot {\mathbf B}^{H})|^2}{\mathrm{tr}(\mathbf B \mathbf R_x \mathbf B^{H})} \ge t\\ \notag
  & \quad  \eqref{eq:power}~\text{and}~ \eqref{eq:semi}.
\end{align}
\end{subequations}
Using the Schur complement, constraint \eqref{eq:Schur’s complement} is equivalently  transformed into the following convex semi-definite constraint:
\begin{equation}\label{eq:SC}
\left[\begin{array}{cc}
      \mathrm{tr}(\dot {\mathbf B} \mathbf R_x \dot {\mathbf B}^{H})-t & \mathrm{tr}(\mathbf B \mathbf R_x  \dot {\mathbf B}^{H})  \\
      \mathrm{tr}(\dot {\mathbf B} \mathbf R_x \mathbf B^{H} ) &  \mathrm{tr}(\mathbf B \mathbf R_x\mathbf B^{H})   
\end{array}\right] \succeq \mathbf{0}.
\end{equation}
Accordingly, problem (P2.1) is equivalent to the following semi-definite program (SDP), which can be optimally solved by convex solvers such as CVX \cite{cvx}:
\begin{subequations}
\begin{align} \notag
  \text{(P2.2)}: \ \ & \max_{\mathbf R_x,t}  \  \  t \\  \notag
  \text{s.t.}&   \ \ \eqref{eq:power}, ~ \eqref{eq:semi}, ~ \text{and}~ \eqref{eq:SC}.
\end{align}
\end{subequations}

\subsection{Reflective Beamforming Optimization}
Next, we optimize the reflecting beamformer $\mathbf v$ in problem (P1) under any given  transmit beamformers $\mathbf R_x$, in which the CRLB formula in \eqref{eq:CRB_1} is used. In this case, the reflective beamforming optimization problem is formulated as
\begin{subequations}
\begin{align} \notag
  \text{(P3)}:  \ \  \max_{\mathbf{v}}  \  \  &\mathbf v^{H} \mathbf R_2 \mathbf v \left(\mathbf v^{H} \mathbf D \mathbf R_1 \mathbf D\mathbf v-\frac{|\mathbf v^{H} \mathbf D \mathbf R_1 \mathbf v|^2}{\mathbf v^{H} \mathbf R_1 \mathbf v}\right)   \\ \notag
  &+  \mathbf v^{H} \mathbf R_1 \mathbf v \left(\mathbf v^{H} \mathbf D \mathbf R_2 \mathbf D \mathbf v-\frac{|\mathbf v^{H} \mathbf D \mathbf R_2 \mathbf v|^2}{\mathbf v^{H} \mathbf R_2 \mathbf v}\right) \\\notag
  \text{s.t.}  \ \  & \ \ \eqref{eq:phase_1},
\end{align}
\end{subequations}
which is still non-convex due to the non-concavity of the objective function and the unit-modulus constraint in \eqref{eq:phase_1}. To resolve this issue, in the following we first deal with constraint \eqref{eq:phase_1} based on the idea of SDR, and then use SCA to approximate the relaxed problem as convex ones. 

First, we use the idea of SDR to deal with the unit-modulus constraints in \eqref{eq:phase_1}.
We define $\mathbf{V}=\mathbf{v} \mathbf{v}^{H}$ with $\mathbf{V} \succeq \mathbf{0}$ and $\mathrm {rank}(\mathbf{V})=1$. As a result, it follows based on \eqref{eq:phase_1} that $\mathbf V_{n,n}=1,\forall n\in \{1,\cdots, N\}$. Also, we have  $\mathbf v^{H} \mathbf R_1 \mathbf v = \mathrm{tr}(\mathbf R_1 \mathbf V)$, $\mathbf v^{H} \mathbf R_2 \mathbf v = \mathrm{tr}(\mathbf R_2 \mathbf V)$, $\mathbf v^{H} \mathbf D \mathbf R_1 \mathbf v = \mathrm{tr}(\mathbf D \mathbf R_1 \mathbf V)$, $\mathbf v^{H} \mathbf D \mathbf R_2 \mathbf v = \mathrm{tr}(\mathbf D \mathbf R_2 \mathbf V)$, $\mathbf v^{H} \mathbf D \mathbf R_1 \mathbf D\mathbf v=\mathrm{tr}(\mathbf D \mathbf R_1 \mathbf D\mathbf V)$, and $\mathbf v^{H} \mathbf D \mathbf R_2 \mathbf D\mathbf v=\mathrm{tr}(\mathbf D \mathbf R_2 \mathbf D\mathbf V)$. 
By substituting $\mathbf{V}=\mathbf{v} \mathbf{v}^{H}$ and introducing two auxiliary variables $t_1$ and $t_2$, problem (P3) is equivalently re-expressed as 
\setcounter{equation}{15} 
\begin{subequations}
\begin{align} \notag
  \text{(P3.1)}:  \max_{\mathbf V, t_1, t_2} \  \  &f_1(\mathbf V,t_1,t_2)+f_2(\mathbf V,t_1,t_2)\\\label{eq:p3_st_1}
  \text{s.t.}  \ \  &\mathbf V_{n,n}=1,\forall n\in \{1, \cdots , N\}\\\label{eq:p3_st_2}
  \qquad & \mathbf{V} \succeq \mathbf{0}\\ \label{eq:rank_1}
  \qquad & \mathrm {rank}(\mathbf{V})=1 \\ \label{eq:p3.3st_1}
  \qquad &t_1\ge \frac{|\mathrm{tr}(\mathbf D\mathbf R_1\mathbf{V})|^2}{\mathrm{tr}(\mathbf R_1\mathbf{V})}\\\label{eq:p3.3st_2}
  \qquad &t_2\ge \frac{|\mathrm{tr}(\mathbf D\mathbf R_2\mathbf{V})|^2}{\mathrm{tr}(\mathbf R_2\mathbf{V})},
\end{align}
\end{subequations}
where $f_1(\mathbf V,t_1,t_2)
=\frac{1}{4}\mathrm{tr}^2((\mathbf R_2+\mathbf D\mathbf R_1\mathbf D)\mathbf{V})+\frac{1}{4}(\mathrm{tr}(\mathbf R_2 \mathbf V-t_1)^2
  +\frac{1}{4}\mathrm{tr}^2((\mathbf R_1+\mathbf D\mathbf R_2\mathbf D)\mathbf{V})+\frac{1}{4}(\mathrm{tr}(\mathbf R_1 \mathbf V-t_2)^2$
 and 
$f_2(\mathbf V,t_1,t_2)
=-\frac{1}{4}\mathrm{tr}^2((\mathbf R_2-\mathbf D\mathbf R_1\mathbf D)\mathbf{V})-\frac{1}{4}(\mathrm{tr}(\mathbf R_2 \mathbf V+t_1)^2-\frac{1}{4}\mathrm{tr}^2((\mathbf R_1-\mathbf D\mathbf R_2\mathbf D)\mathbf{V})-\frac{1}{4}(\mathrm{tr}(\mathbf R_1 \mathbf V+t_2)^2$. Here, $f_1(\mathbf V,t_1,t_2)$ and  $f_2(\mathbf V,t_1,t_2)$ are convex and concave functions, respectively.
Using the Schur complement, constraints \eqref{eq:p3.3st_1} and \eqref{eq:p3.3st_2} are equivalently transformed into the following convex semi-definite constraints:
\begin{equation}\label{eq:sc_2}
   \left[\begin{array}{cc}
      t_1 & \mathrm{tr}(\mathbf D\mathbf R_1 \mathbf V)  \\
      \mathrm{tr}(\mathbf V^{H}\mathbf R_1^{H} \mathbf D^{H}) &  \mathrm{tr}(\mathbf R_1 \mathbf V)   
\end{array}\right] \succeq \mathbf{0},
\end{equation}
\begin{equation}\label{eq:sc_3}
\left[\begin{array}{cc}
      t_2 & \mathrm{tr}(\mathbf D\mathbf R_2 \mathbf V)  \\
      \mathrm{tr}(\mathbf V^{H}\mathbf R_2^{H} \mathbf D^{H}) &  \mathrm{tr}(\mathbf R_2 \mathbf V)   
\end{array}\right] \succeq \mathbf{0}.
\end{equation}

Problem (P3.1) is then equivalent to the following problem:
\begin{subequations}
\begin{align} \notag
  \text{(P3.2)}:  \max_{\mathbf V, t_1, t_2} & \  \  f_1(\mathbf V,t_1,t_2)+f_2(\mathbf V,t_1,t_2)\\\notag
  \text{s.t.}  \ \  &\eqref{eq:p3_st_1},~\eqref{eq:p3_st_2},~\eqref{eq:rank_1},~\eqref{eq:sc_2},~\text{and}~\eqref{eq:sc_3}.
\end{align}
\end{subequations}
By dropping the rank-one constraint in \eqref{eq:rank_1}, we get the relaxed version of problem (P3.2) without constraint \eqref{eq:rank_1}, denoted by problem (SDR3.2). Note that in problem (SDR3.2), all constraints are convex and only $f_1(\mathbf V,t_1,t_2)$ in the objective function is non-concave. 

To solve the non-convex problem (SDR3.2), we use SCA to approximate it as a series of convex problems. The SCA-based solution to problem (SDR3.2) is implemented in an iterative manner as follows. Consider each  inner iteration $r \ge 1$, in which the local point is denoted by $\mathbf V^{(r)}$, $t_1^{(r)}$, and $t_2^{(r)}$. Then based on the local point, we obtain a global linear lower bound function $\hat f_1^{(r)}(\mathbf V,t_1,t_2)$ for the convex function $f_1(\mathbf V,t_1,t_2)$ in problem (SDR3.2), based on its first-order Taylor expansion.

Replacing $f_1(\mathbf V,t_1,t_2)$ by $\hat f_1^{(r)}(\mathbf V,t_1,t_2)$, problem (SDR3.2) is approximated as the following convex form in inner iteration $r$:
\begin{subequations}
\begin{align}\notag 
 \text{(P3.3.}r\text{)}: \max_{\mathbf V,t_1,t_2}&  \quad  \hat f_1^{(r)}(\mathbf V,t_1,t_2)+f_2(\mathbf V,t_1,t_2)\\\notag
  \text{s.t.} & \quad  \eqref{eq:p3_st_1},~\eqref{eq:p3_st_2},~ \eqref{eq:sc_2},~\text{and}~ \eqref{eq:sc_3},
\end{align}
\end{subequations}
which can be optimally solved  by convex solvers such as CVX. Let $\hat{\mathbf V}^{\star}$, $\hat t_1^{\star}$, and $\hat t_2^{\star}$ denote the optimal solution to problem (P3.3.$r$), which is then updated to be the local point $\mathbf V^{(r+1)}$, $t_1^{(r+1)}$, and $t_2^{(r+1)}$ for the next inner iteration. Since $\hat f_1^{(r)}(\mathbf V,t_1,t_2)$ serves as a lower bound of $f_1(\mathbf V,t_1,t_2)$, it is ensured that $f_1(\mathbf V^{(r+1)}, t_1^{(r+1)}, t_2^{(r+1)}) + f_2(\mathbf V^{(r+1)}, t_1^{(r+1)}, t_2^{(r+1)})\ge f_1(\mathbf V^{(r)}, t_1^{(r)}, t_2^{(r)})+ f_2(\mathbf V^{(r)}, t_1^{(r)}, t_2^{(r)})$, i.e., the inner iteration leads to a non-decreasing objective value for problem (SDR3.2). Therefore, the convergence of SCA for solving problem (SDR3.2) is ensured. Let $\tilde{\mathbf V}$, $\tilde t_1$, and $\tilde t_2$ denote the obtained solution to problem (SDR3.2) based on SCA, where $\mathrm{rank}(\tilde {\mathbf V}) >1$ may hold in general.

Finally, we construct an approximate rank-one solution $\mathbf V$ to problem (P3.2) or (P3) using Gaussian randomization. Specifically, we first generate a number of random realizations $\mathbf z \sim \mathcal{CN}(\mathbf{0},\tilde{\mathbf V})$, and accordingly construct a set of candidate feasible solutions as $\mathbf{v}=e^{j\mathrm {arg}(\mathbf z)}$. We then choose the $\mathbf v$ that achieves the maximum objective value of problem (P3.2) or (P3). 
%Note that the Gaussian randomization should be implemented sufficiently many times to ensure that the objective value increases at each outer iteration of alternating optimization.

%\subsection{Complete Algorithm of Alternating Optimization}

%Based on the results in Sections IV-A and IV-B, the alternating optimization based algorithm for solving problem (P1) is complete. 
%In each outer iteration, we first solve problem (P2) to update the transmit beamformers $\mathbf R_x$ with the updated reflective beamformer $\mathbf v$, and then solve problem (P3) to update $\mathbf v$ with the updated $\mathbf R_x$. 
In each outer iteration of alternating optimization, problem (P2) is optimally solved, thus leading to a non-increasing CRLB value. Also notice that with sufficient number of Gaussian randomizations, the obtained solution to problem (P3) is also ensured to result in a monotonically non-increasing CRLB. As a result, the convergence of the proposed algorithm for solving problem (P1) is ensured.

\section{Numerical Results}
This section provides numerical results to evaluate the performance of our proposed joint beamforming design based on CRLB minimization. We consider the Rician fading channel for the AP-IRS link with the Rician factor being $0.5$. We consider the distance-dependent path loss as $K_0(\frac{d}{d_0})^{-\alpha_0}$, where $d$ is the distance of the transmission link and $K_0=-30~ \text{dB}$ is the path loss at the reference distance $d_0=1~ \text{m}$, and the path-loss exponent $\alpha_0$ is set as $2.5$ for the AP-IRS and IRS-target links. The AP, target, and IRS are located at coordinate $(0,0)$, $(5~\text{m},0)$, and $(5~\text{m},5~\text{m})$, respectively. We set the number of antennas at the AP, the number of reflecting elements at the IRS, and the radar dwell time slots as $M = 8$, $N = 8$, and $T=256$, respectively. We also set the target's RCS as one and the noise power at the AP as $\sigma_\text{R}^2 = -120~ \text{dBm}$, respectively.

\begin{figure}[t]
    \centering
    \includegraphics[width=0.3\textwidth]{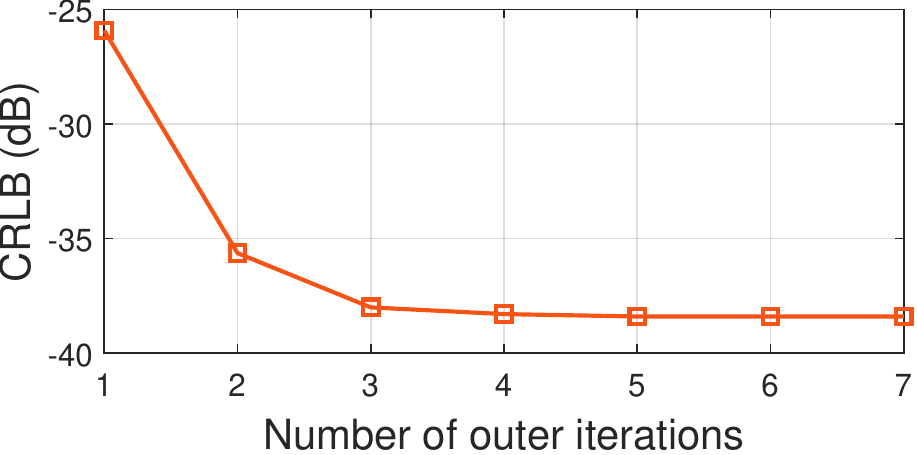}
    \caption{Convergence behavior of the proposed alternating optimization based algorithm for solving problem $\text{(P1)}$, where $P_0 = 30~ \text{dBm}$.}
    \label{convergence}
\end{figure}

Fig.~\ref{convergence} shows the convergence behavior of our proposed alternating optimization based algorithm for solving problem $\text{(P1)}$, where $P_0 = 30~ \text{dBm}$. It is observed that the proposed alternating optimization based algorithm converges within around $5$ outer iterations, thus validating its effectiveness.

Next, we evaluate the estimation performance of our proposed joint beamforming design as compared to the following benchmark schemes. 

\subsubsection{SNR maximization} We maximize the received SNR of the echo signal at the AP in \eqref{eq:echo_signal}, by jointly optimizing the transmit beamforming at the AP and reflective beamforming at the IRS. It can be shown that maximum-ratio transmission is the optimal transmit beamforming solution. The reflective beamformer $\mathbf v$ at the IRS is optimized to maximize the channel norm of the AP-IRS-target link $\|\mathbf G^{T} \mathbf A \mathbf v\|^2$, subject to the unit-modulus constraint in \eqref{eq:phase_1}, which is similar to the SNR maximization in IRS-enabled multiple-input single-output (MISO) communications that has been studied in \cite{8811733}.

\subsubsection{Reflective beamforming only with isotropic transmission (Reflective BF only)} The AP uses the isotropic transmission by transmitting orthonormal signal beams and setting $\mathbf R_x = P_0/M\mathbf I_M$. Then, the reflective beamforming at the IRS is optimized to minimize the CRLB by solving problem (P3).

\subsubsection{Transmit beamforming only with random phase shifts (Transmit BF only)} We consider the random reflecting phase shifts at the IRS. Then,  the transmit beamforming at the AP is optimized to minimize the CRLB by solving problem (P2).

Furthermore, besides CRLB, we also implement the practical MLE method to estimate the target's DoA $\theta$, and accordingly evaluate the estimation MSE as the performance metric for gaining more insights. Refer to the extended version of this paper \cite[Appendix E]{xianxin} for the details of MLE for DoA estimation.

\begin{figure}[t]
    \centering
    \includegraphics[width=0.4\textwidth]{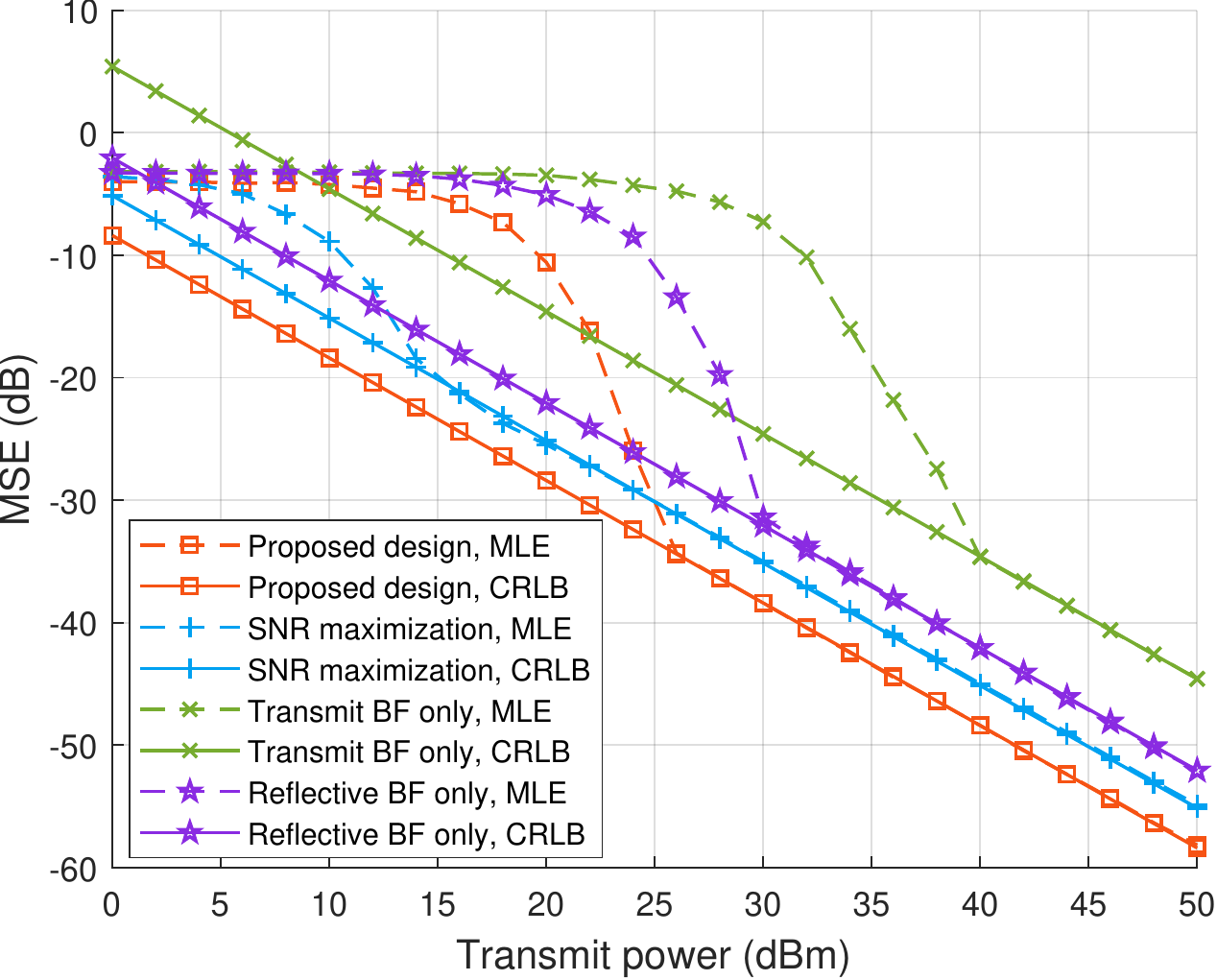}
    \caption{The MSE performance for estimating DoA versus the transmit power $P_0$ at the AP.}
    \label{correctness}
\end{figure}

Fig.~\ref{correctness} shows the CRLB and the MSE with MLE versus the transmit power $P_0$ at the AP. It is observed that the CRLB in decibels (dB) is monotonically decreasing in a linear manner w.r.t. the transmit power $P_0$ in dB at the AP. This is because that the $\mathrm{CRLB}(\theta)$ in \eqref{eq:CRB} is inversely proportional to the transmit power $P_0$. 
Furthermore, for each of the four schemes, the estimation MSE achieved by MLE is observed to be identical to the derived CRLB at the high SNR regime, but deviate from the CRLB when the SNR is below a certain threshold (e.g., $P_0 = 24 ~ \text{dBm}$ for the proposed design). In addition, when the SNR becomes sufficiently low, the estimation MSE by MLE is observed to converge to a constant value. This is due to the fact that the estimated DoA with MLE must be within the interval of $[-\frac{\pi}{2}, \frac{\pi}{2}]$, thus making the estimation error bounded. These observations are consistent with the results in \cite{kay1993fundamentals,1624646} for DoA estimation without IRS, thus validating the correctness of our CRLB derivation.
It is also observed that the proposed CRLB minimization scheme achieves the lowest CRLB in the entire transmit power regime, which shows the effectiveness of our proposed joint beamforming design.
Furthermore, when the transmit power is sufficiently high (e.g., $P_0 > 25 ~ \text{dBm}$), CRLB minimization achieves lower MSE than the three benchmark schemes. When the transmit power is low (e.g., $P_0 < 25 ~ \text{dBm}$), SNR maximization achieves lower MSE (with MLE) than the proposed CRLB minimization scheme. This is because in this case, the CRLB is not achievable using MLE, and the estimation performance of MLE is particularly sensitive to the power of echo signals, thus making the SNR maximization scheme desirable.

\section{Conclusion}
This work considered an IRS-enabled NLoS wireless sensing system consisting of an AP with multiple antennas, an ULA-IRS with multiple reflecting elements, and a target in the NLoS region of the AP. The AP aimed to estimate the target's DoA w.r.t. the IRS based on the echo signal received at the AP from the AP-IRS-target-IRS-AP link. Under this setup, we jointly design the transmit beamforming at the AP and the reflective beamforming at the IRS to minimize the DoA estimation error in terms of CRLB. Towards this end, we obtained the CRLB for DoA estimation in closed form and then minimized the obtained CRLB by jointly optimizing the transmit and reflective beamforming. Numerical results showed that the proposed algorithm achieved improved estimation performance in terms of estimation MSE, especially when the SNR became large. 
%We envision that this work can provide new insights in designing IRS-enabled sensing and IRS-enabled ISAC systems.  

\ifCLASSOPTIONcaptionsoff
  \newpage
\fi

\bibliographystyle{IEEEtran}
\bibliography{IEEEabrv,mybibfile}

\end{document}